\documentclass[conference]{IEEEtran}
\IEEEoverridecommandlockouts
\usepackage{cite}
\usepackage{amsmath,amssymb,amsfonts}
\usepackage{algorithmic}
\usepackage{graphicx}
\usepackage{multirow}
\usepackage{textcomp}
\usepackage{xcolor}
\def\BibTeX{{\rm B\kern-.05em{\sc i\kern-.025em b}\kern-.08em
    T\kern-.1667em\lower.7ex\hbox{E}\kern-.125emX}}

\begin{document}

\title{Bank transactions embeddings help to uncover current macroeconomics\\
% {\footnotesize \textsuperscript{*}Note: Sub-titles are not captured in Xplore and
% should not be used}
% \thanks{Identify applicable funding agency here. If none, delete this.}
}

\author{\IEEEauthorblockN{1\textsuperscript{st} Maria Begicheva}
\IEEEauthorblockA{\textit{CDISE} \\
\textit{Skolkovo Institute of Science and Technology}\\
Moscow, Russia \\
maria.begicheva@skoltech.ru}\\
\and
% \IEEEauthorblockN{2\textsuperscript{nd} Oleg Travkin}
% \IEEEauthorblockA{\textit{Department of Integrated Risk Management} \\
% \textit{Sberbank}\\
% Moscow, Russia \\
% travkin.o.i@gmail.com}
% \and
\IEEEauthorblockN{2\textsuperscript{nd} Alexey Zaytsev}
\IEEEauthorblockA{\textit{CDISE} \\
\textit{Skolkovo Institute of Science and Technology}\\
Moscow, Russia \\
a.zaytsev@skoltech.ru}\\
}

% \IEEEauthorblockN{4\textsuperscript{th} Given Name Surname}
% \IEEEauthorblockA{\textit{dept. name of organization (of Aff.)} \\
% \textit{name of organization (of Aff.)}\\
% City, Country \\
% email address}
% \and
% \IEEEauthorblockN{5\textsuperscript{th} Given Name Surname}
% \IEEEauthorblockA{\textit{dept. name of organization (of Aff.)} \\
% \textit{name of organization (of Aff.)}\\
% City, Country \\
% email address}
% \and
% \IEEEauthorblockN{6\textsuperscript{th} Given Name Surname}
% \IEEEauthorblockA{\textit{dept. name of organization (of Aff.)} \\
% \textit{name of organization (of Aff.)}\\
% City, Country \\
% email address}

\maketitle
% \IEEEpeerreviewmaketitle

\begin{abstract}
Macroeconomic indexes are of high importance for banks: many risk-control decisions utilize these indexes.
A typical workflow of these indexes evaluation is costly and protracted, with a lag between the actual date and available index being a couple of months.
Banks predict such indexes now using autoregressive models to make decisions in a rapidly changing environment. However, autoregressive models fail in complex scenarios related to appearances of crises.

We propose to use clients' financial transactions data from a large Russian bank to get such indexes. Financial transactions are long, and a number of clients is huge, so we develop an efficient approach that allows fast and accurate estimation of macroeconomic indexes based on a stream of transactions consisting of millions of transactions. 
The approach uses a neural networks paradigm and a smart sampling scheme.

The results show that our neural network approach outperforms the baseline method on hand-crafted features based on transactions. 
Calculated embeddings show the correlation between the client's transaction activity and bank macroeconomic indexes over time.

\end{abstract}


\section{Introduction}
The coronavirus pandemic has caused great damage to the health and changed lives of many millions of people around the world. As the state tried its best to stop the further spread of the virus or at least slow it down, their actions caused economic problems. Many industries suffered, affecting the banking industry. 
In particular, machine learning models degraded due to data distribution shift and other reasons~\cite{covid19}. 
Models for assessing the loan portfolio also suffered, as they always do during crises~\cite{sousa2014two,sousa2016new}.
The shortcomings of these models surfaced and the question arose whether it was worth trusting them at all, if they provide little help during vital crisis times for banks.

% There are several reasons why machine learning models stopped working. The first reason is that the assumptions and constraints inherent in the model were developed in the world before the crisis. The second reason is that these models use historical data as a data source, which does not allow the model to be calibrated in accordance with the current macroeconomics. Well, the third reason is that although there is access to an alternative data source, it is not so easy to integrate new information into the model since the model needs to be completely rebuilt for this.

% As a result, since some models fail, the models which use the predictions of the previous ones also stopped working. Banks need to revise their models and include the ability to calibrate, that's a fact. Models need to be revisited, both short-term and long-term. This approach is good in that by adjusting the model in the short-term horizon, the bank ensures the continuity of the flow of business processes, while at the same time it is possible to analyze how to correct the model in the long-term horizon.

Existing credit risk models in bank are of two types: one type describes client behaviour at the application stage and following lifetime of this particular credit, second type describes lifetime of the whole portfolio. Models of the first type use the loan application form and the client's credit history as data. These models use machine learning techniques such as logistic regression or decision trees to estimate a customer's probability of default (PD), which is the probability of repaying a debt to the bank. Moreover, novel models use deep learning methods, for example, Embedding-Transactional Recurrent Neural Network (E.T.-RNN) rely on history of bank customers' credit and debit card transactions to compute PDs~\cite{babaev2019rnn}.

Both types of models suffer from the change of macroeconomics. 
For e.g. some borrower is identified as a "good" one by model, but when COVID-19 happened this borrower defaulted on a debt, because he lost his job and was unable to pay his payments. 
Second type of models are linear multivariate autoregressive models, that forecast several different important indexes for bank. 
There is no correction for macroeconomics in these models also. 

\begin{figure}
    \centering
    \includegraphics[width=\columnwidth]{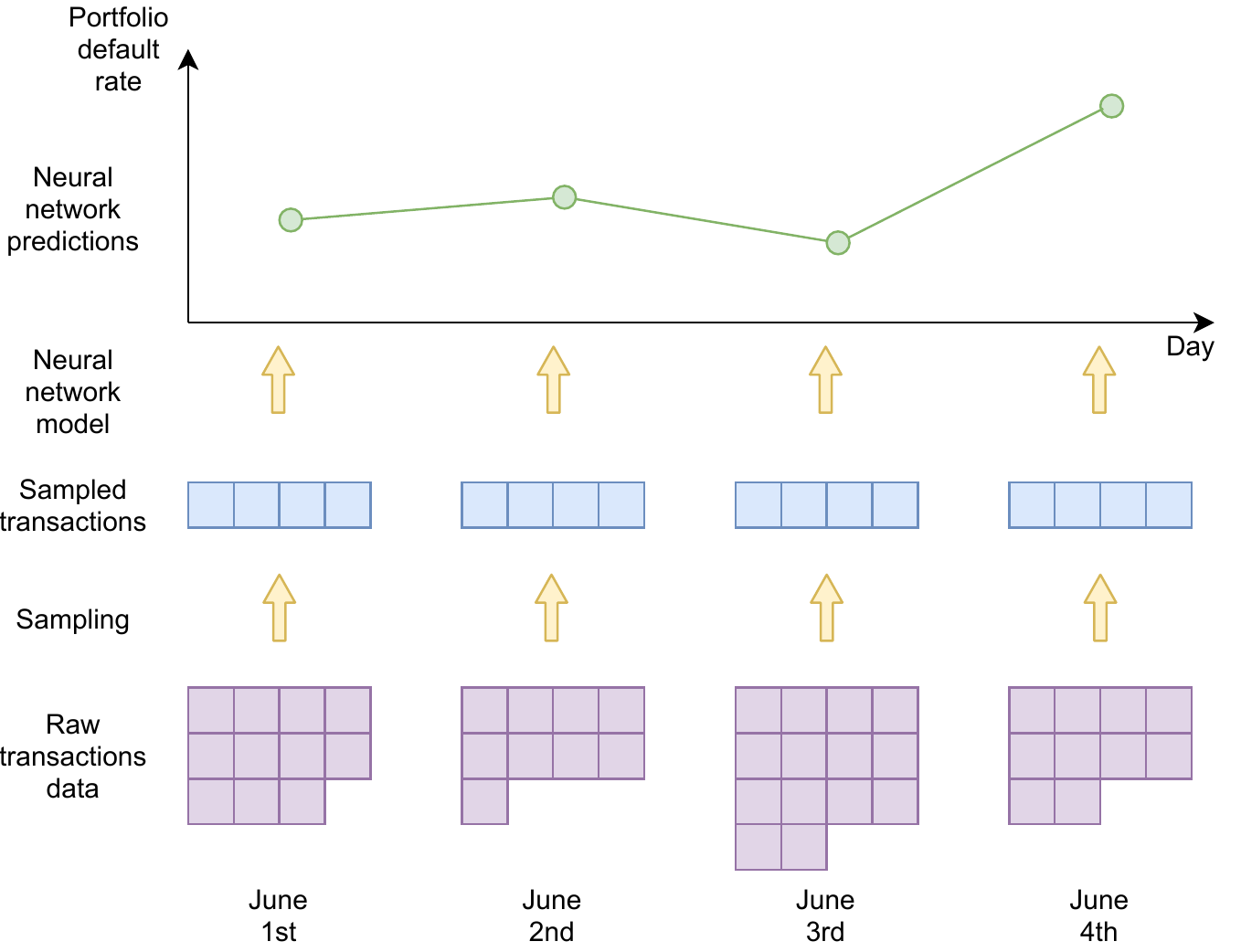}
    \caption{Transactions processing scheme for our approach: from an ocean of available client financial transactions made during a particular day we sample a useful part, then we process them via a neural network to produce a prediction of a macroeconomic indexes for these day. The scheme allows end2end learning of neural network parameters to directly predict the target, which is the portfolio default rate in our case.}
    \label{fig:general_scheme}
\end{figure}

In the current work we consider an important bank's macroeconomic index: a default rate. 
The default rate is a fraction of bank clients, who went overdue 90+ days for the current date~\cite{cotugno2013relationship}.
The main goal of this work is to show the correlation between people transaction activity and default rate over time.
The scheme of our approach is presented in Figure~\ref{fig:general_scheme}.

Our main claims are the following:
\begin{itemize}
    \item We proposed an approach that can work with the data of complex structure: we have a time series, and each object in this time series is a long sequence of events.
    \item Our neural network approach can use long sequences of events with hundreds of thousands of events.
    \item We apply our approach to the problem of macroeconomic indexes. We predict the daily portfolio default rate using financial transactions events of bank clients.
    \item Our models work better than alternatives including classic machine learning models with hand-crafted features.
    \item Obtained day embeddings are meaningful, and thus can be used to solve other downstream tasks.
\end{itemize}

\section{Literature review}

Our problem has two essential components: the problem of macroeconomic indexes prediction without specification of particular input data and the problem of representation learning for structure data, in particular usage of long sequences of events as input to a deep learning model. In the review we separate these two directions.
% Papers discussing the stated problem can be divided into two parts: macroeconomic indexes . From first point of view, the problem could be addressed as consideration of macroeconomics in the area of economics field. From the second point of view, the stated topic could be investigated as an embedding calculation task in machine Learning and deep Learning. The Literature review is organized as follows: consideration of macroeconomics in the economics field, general model types in embedding calculation.

\textit{Economics} papers consider prediction of macroeconomic indexes using external data, in particular transactions aggregates. In \cite{gil2018nowcasting}
authors focus on now-casting and forecasting quarterly private consumption. The authors used several external data sources, in particular the data collected from automated teller machines (ATMs), encompassing cash withdrawals at ATM terminals, and points-of-sale (POS) payments with debit and credit cards. The received data is transformed into aggregates of transactions (amount feature). In addition, the considered data included the internet search patterns provided by Google Trends to construct indicators of consumption behavior, traditional monthly indicators as employment growth, index of retail sales and other. As a benchmark authors used random-walk model and compare it with autoregressive models (AR) built on different indicators. Detailed analysis demonstrated that transaction indicators and traditional indexes seem to dominate the rest indicators. Moreover, the model that used transaction aggregates and traditional indexes as inputs is better, than the model that used only indexes in terms of RMSE for different prediction horizon.
In \cite{article2} the authors used the debit card data to predict quarterly Norwegian total household consumption. Aggregated debit card data come from the Norwegian retail clearing institution and are sum of the values of all transactions within a week.
As a benchmark authors used autoregressive model built on past values of the target variable and compare it with AR models built on other quarter predictors and with mixed-data sampling (MIDAS) regressions built on available weekly data~\cite{article3}. MIDAS regressions with debit card transactions data improve predicted accuracy compare to standard AR models that use alternative quarter predictors over the sample 2011Q4-2020Q1.
In \cite{article4} the authors construct a graph of financial transactions, where the edges are described with the following data: transaction amount and frequency, which characterize the committed transaction. The nodes in these graphs contain information about account balance. The authors focused on credit card transactions, which form a bipartite graph between holders of credit cards and the merchants where they shop at. The main challenge is that these graphs are highly dynamic. Authors applied Graph Neural Networks to encode high-dimensional data in low-dimensional latent space. Calculating shifts of embeddings in the latent space lead to quantitative estimation of impact of economic events (in particular COVID-19) happening between these two dates where shifts are measured.
 
\textit{Representation learning} is a major research area in  machine learning and deep learning and there are a lot of articles devoted to it. The goal is to construct small-dimensional representations of complex structured data~\cite{goodfellow2016deep}, e.g. images~\cite{weiss2016survey} or sequences of events~\cite{zhuzhel2021cohortney}. 
We will start from simple approach moving toward more complicated ones.

In \cite{simple} the authors suggest sentence embedding method, improving simple averaging of words in a sentence. The algorithm has two steps: (1) calculate the weighted average of the words' embeddings in the sentence, (2) remove the projections of the average vectors on their first singular vector.
The weight of each word $w$ is $a/(a + p(w))$, with $a$ being a hyperparameter of the algorithm and $p(w)$ is the word frequency, which can be estimated in several ways. For text similarity problems this method shows better results than the simple average of the vectors of words in a sentence, and even surpasses RNN and LSTM models in most problems. 
This method is easy to implement, unsupervised and can work with no limits in length of sequence lengths.
Also one can apply this method to the calculation embeddings from transaction data. Here a sequence of transactions is a sentence, and a single transaction is a word from a finite dictionary. The word frequency $p(w)$ is the frequency of specific Merchant Category Code (MCC-code) for transaction in a dictionary.

In \cite{article5} the authors develops a model for the construction of sentence embedding using Recurrent Neural Network (RNN) \cite{Sherstinsky_2020} with Long Short-Term Memory (LSTM) cells. The proposed model creates semantic vector from sentence by sequentially processing each word, extracting information that contains in it and union all information from all words. The model accumulates a large amount of information to the last word in sentence due to LSTM sells, integrated in model, which allow to capture long term memory. To the last word hidden layer of this network contains information about semantic meaning of the sentence.
RNNs have been used to process sequential data and compute their embeddings. It worked well for quite short sentences. But when the sequence length increases, the difficulty in training this network arises due to gradient vanishing and exploding problems~\cite{kag2019rnns}. For RNNs learning long-term patterns is quite difficult. To tackle this problems, new architectures were developed: Long Short-Term Memory (LSTM)~\cite{hochreiter1997long} and Gated Recurrent Unit (GRU)~\cite{cho2014learning}. However, the length of sequences on which these networks could be trained is still limited by about 1000 tokens or "words" due to gradient decay over layers.
% (results of applying hyperbolic tangent and the sigmoid functions). 
%Also the neurons in RNN all connected together, which causing the problem with interpretability. 
That's why constructing a good trainable neural network for long sequences is a challenging task.

In~\cite{indrnn} the authors claim that the problems with RNN and LSTM described above can be dealed with, if we change the architecture of RNN a bit. The proposed new architecture is called Independent Recurrent Neural Network (IndRNN). The difference between this architecture from standard RNN is in independence of neurons in the same layer, but it doesn't mean that there is no connection between neurons at all: neurons are connected only across layers. 

The authors have shown that the IndRNN network can easily be trained on long sequences, it is not exposed to gradient exploding and vanishing problems. In addition, this network can be trained robustly with ReLU (Rectified Linear Unit) activation function. To construct deeper neural network than existing RNNs, authors propose to stack Multiple IndRNNs into one big model. Experiments showed that proposed model with 21 layers can be trained robustly on sequences of length over 5000. Comparison of IndRNNs with RNN and LSTM on various problems showed superiority of proposed IndRNNs. Better performances have been achieved on various tasks by using IndRNNs compared with the traditional RNN and LSTM.

In \cite{cnn} the authors suggest to use Convolution Neural Networks (CNN) to calculate embeddings via solving time series classification problem, which is quite challenging due to the specific nature of time series data: high feature dimensionality, big data size and noisy nature of data itself. Traditionally feature-based approaches were used to tackle this problem, but it's obvious that generating hand-crafted features from time-series leads to reducing the original information. The deep learning techniques are explored to improve the performance of traditional feature-based approaches. Convolution neural network framework is proposed for this classification problem in order to improve the performance of traditional methods.
The difference from other traditional approaches is in ability of CNN to automatically extract features from time series data by using convolution and pooling operations which fully describes the structure of the data. The experimental results for variety of tasks in time series classification show that the proposed model outperforms state-of-the-art methods in terms of the classification accuracy and noise tolerance. The main advantage of CNNs compared to RNN and their derivatives is that it can be applied to very long sequences (sequences of length that will fit GPU memory and it's about hundreds of thousands). The problem is in such long sequences sequential nature of data will be lost yielding more pattern-like nature to step forward.

One more approach that should be mentioned is using transformers for embedding learning \cite{article6}. In this work authors present method for unsupervised embedding learning from unlabeled data. A new state-of-the-art unsupervised method based on pre-trained Transformers and Sequential Denoising Auto-Encoder (TSDAE) which outperforms previous approaches by up to 6.4 points. It can achieve up to 93.1\% of the performance of in-domain supervised approaches. Further, authors show that TSDAE is a strong pre-training method for learning sentence embeddings, significantly outperforming other approaches like Masked Language Model.
A crucial shortcoming of previous studies is the narrow evaluation: most work mainly evaluates on the single task of Semantic Textual Similarity (STS), which does not require any domain knowledge. It is unclear if these proposed methods generalize to other domains and tasks. Authors fill this gap and evaluate TSDAE and other recent approaches on four different datasets from heterogeneous domains. The main disadvantage of using transformers in their inability to process long sequences due to their self-attention operation, which scales quadratically with the sequence length.
So, straightforward application of neural networks to long sequences is challenging with even more difficulties for architectures like transformers~\cite{tay2020long}.

\textit{We can conclude,} that the supervised prediction of macro indexes by an end2end approach based on raw financial transaction data is absent in the literature. 
The main challenge here is how should we construct embeddings for such long sequences. 
We propose to conduct a research to answer these questions and evaluate feasibility of such ideas.

\section{Methods}

\subsection{Data structure}
We consider two datasets: the one available to the bank on prediction of daily credit default rate for a portfolio of clients and the open one related to the prediction of a weekday for financial transactions.
The overall characteristics of datasets are in Table~\ref{table:data}.

During numerical experiments we split the data, so the training sample includes 80\% of the data, and the remaining 20\% is for the validation. 
The validation data are later in time, than the training data.

\begin{table}[htbp]
\begin{center}
\begin{tabular}{ccc}
\hline
 
\textbf{Feature} & \textbf{Day of Week} & \textbf{Default rate} \\
\hline
Target & 7 classes & Float \\
MCC-code & 184 & 396  \\
Transaction type & 77 & 134 \\
Currency & - & 111 \\
Country & - & 303 \\
Time & \multicolumn{2}{c}{Number of hours from 00:00} \\
Amount & \multicolumn{2}{c}{Float} \\
\hline
\end{tabular}
\caption{Features of datasets marked by their target variable. For categorical variable we provide number of unique values presented in the corresponding dataset.}
\label{table:data}
\end{center}
\end{table}

\paragraph{Day-of-Week data.} To provide wider understanding of strengths and weaknesses of our methods we as well consider open transactions data on prediction of a day-of-week on the base of clients' transactions.

% The open dataset consists of 3 files: first table contains the history of transactions of bank clients for one year and three months, second table contains the description of MCC-codes of transactions, third table contains the description of the types of transactions. 
The open dataset~\cite{fursov2021adversarial} consists of the history of transactions of bank clients for one year and three months.
Information about day of week for these transactions is available for each day. 
Number of transactions in dataset is 7 millions, with maximum number of transactions per day 20500, minimal number is 5600, mean number is 15000. The data include 4 input features for each transaction and target labels presented in the above table. 

Our idea was to check if we can predict day of week by transactions committed during this day.

\paragraph{Default rate data.} The source of data for current research is a large Russian bank. The data consists of more than $1$ billion transactions records for $2$ years and overall portfolio default rate for each day from this period. From this dataset $110$ millions of transactions were uniformly randomly selected for the research to speed up experiments, as further dataset increase doesn't lead to the model improvement. 
For each day we have about 130 thousands of transactions, making straightforward application of deep learning methods impossible.
The data includes 6 input features for each transaction and the target portfolio default rate labels.

% The training data consist of data from period starting from January 2018 to September 2019.
% The test data consist of data from period starting from October 2019 to April 2020. 

\subsection{Predictive model}

There are tremendous amount of available data, so we should deal with them in an effective manner.  

% The global idea is to use a neural network architecture for the processing of long transactions sequences. 
% By extracting useful patterns from long sequential data, we will construct embedding, which can be used either in classification task or regression task. 

The idea is to use a standard neural network architecture for the processing of long transactions sequences and apply smart sampling technique during training, which is described in section below. This method will allow to operate all amount of available data. Neural network here play a role of feature extractor: by ejecting useful patterns from long sequential data, we will construct embedding, which can be used either in classification task or regression task.

Specific details of experiments, including neural network parameters are providen in the corresponding section.

\subsection{Training}

\begin{figure}[htbp]
\centerline{\includegraphics[width=1\linewidth]{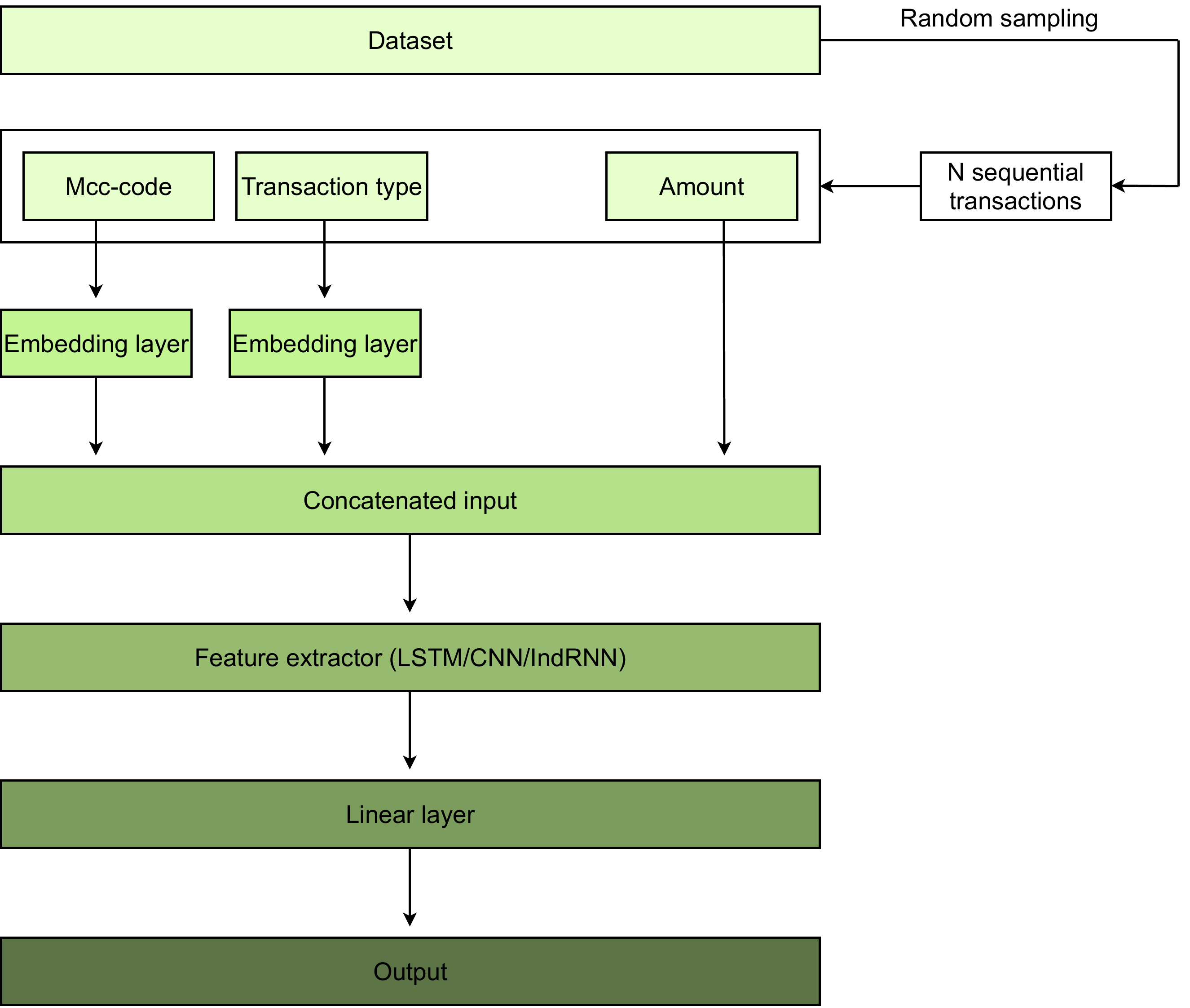}}
\caption{Proposed scheme for the training of a neural network model}
\label{fig}
\end{figure}

The proposed training scheme is in Figure~\ref{fig}.
As we can't use all transactions during an epoch, as they don't fit into available memory, we sample a sequence of uniformly random $N$ different sequential transactions.
Thus, during the whole training we'll use the whole dataset.
The training algorithm has the following steps:
\begin{enumerate}
\item Sample a sequence of uniformly random $N$ different sequential transactions for each day from the whole dataset.
\item Use embedding layers with learnable weights for categorical variables. These layers map every token of its type (MCC-code/transaction type/country/currency) to a vector space, with embedding dimension as a parameter.
\item Concatenate $N$ tokens into one sequence.
\item The resulting sequences serves as the input to a neural network (LSTM/CNN/IndRNN), which output is followed with a pooling layer and a fully connected layer.
\end{enumerate}

\subsection{Inference}

During the inference procedure for the classification task we sample transactions several times from dataset, feed it to neural network, average embeddings and make predictions on obtained embeddings.

During the inference procedure for the regression task we make predictions on one subset of transactions and then apply moving average on predicted values.

\section{Experiments}

\subsection{Methods details}

\paragraph{Baseline.} 
For every calendar day we created handcrafted statistical features from transaction data. We calculated frequencies of each MCC-code, type, country and currency types per day and features, characterizing amount distribution. These features were obtained by grouping amount distribution into 10 bins and calculating mean value of every bin and and number of observations in it. So each day was represented with the 965 tabular features described above.
Via XGBoost library~\cite{chen2016xgboost} we train a gradient boosting with these features as input.
The best hyperparameters found via grid search for XGBoost are the following: number of trees in ensemble is 50, depth of tree is 3, minimum sum of instance weight (Hessian) needed in a child is 1.

\paragraph{Word2vec baseline.} 
We used raw transaction data with preprocessed categorical variables (MCC-code, transaction type) via Word2vec~\cite{mikolov2013efficient}. A good baseline sequence embedding method~\cite{simple} provided unsupervised high-dimensional embeddings. 
On top of these embeddings we use gradient boosting algorithm to predict the target~\cite{chen2016xgboost}.

In our case, word frequency is the frequency of MCC-code. 
The value of hyperparameter in the method was selected according to best metric obtained with boosting model on computed embeddings. Influence of $a$ parameter on embedding space was additionally investigated. The best hyperparameters found via grid search for XGBoost are the following: number of trees in ensemble is 75, depth of tree is 4, minimum sum of instance weight (Hessian) needed in a child is 1.

\paragraph{CNN.} 
We used raw transaction data with preprocessed categorical variables (MCC-code, transaction type) via embedding layer with learnable weights. Experiments were conducted with sequences of lengths from 500 to 5000 with a step of 500. Every epoch we resample sequence with the length of the specified above from dataset. Train includes 80\% of data, and 20\% is for the validation. During validation procedure we sample sequences for 30 times from dataset, return embeddings and evaluate model accuracy on the average of the embeddings. The number of training epochs is $500$, initial learning rate is $\alpha = 0.01$ with step decay coefficient of 0.95 and step size of 5. The following number of training epochs was chosen to provide the best result on validation set. We use Adam optimizer~\cite{kingma2014adam}. Weights are initialized randomly. Embedding sizes are following: $77$, $25$ for MCC-code and transaction type respectively. The batch size is $10$.

\paragraph{IndRNN.}
The difference between IndRNN from standard RNN in independence of neurons in the same layer, but it doesn't mean that there is no connection between neurons at all: neurons are connected only across layers. \\
Experiments were conducted with sequences of the following lengths: 3000, 5000, 10000. Every epoch we resample sequence with a specified length from the dataset. The number of training epochs is $1000$, initial learning rate is $\alpha = 0.001$ with step decay coefficient of 0.95 and step size of 5. The following number of training epochs was chosen to provide the best result on validation set. We use standard Adam optimizer. Weights are initialized randomly. Embedding sizes are following $36$, $18$, $16$, $16$ for MCC-code, transaction type, country and currency respectively. Hidden size is $100$. Batch size is $4$.

\paragraph{LSTM.} We used raw transaction data with preprocessed categorical variables MCC-code and transaction type via an embedding layer with learnable embeddings matrix. 
We consider two different types of loss-functions: a first standard cross-entropy loss and a second triplet loss \cite{triplet} for directly optimizing embeddings.
Experiments were conducted with sequences of lengths from 500 to 1000. Every epoch we resampled sequence with the length of the specified above from dataset. During validation procedure we sample sequences for 30 times from dataset, return embeddings and evaluate model accuracy on the average of the embeddings. The number of training epochs is $100$, initial learning rate is $\alpha = 0.0001$. The following number of training epochs was chosen to provide the best result on validation set. We use standard Adam optimizer. Weights are initialized randomly. Embedding sizes are following: $77$, $25$ for MCC-code and type respectively. Hidden size is $100$. Batch size is $15$. 

\paragraph{t-SNE}
One of our main goals is a construction of useful data representations. 
To examine our progress toward this goal, we embed our representations via T-distributed Stochastic Neighbor Embedding t-SNE~\cite{van2008visualizing}.
We plot 2-dimensional t-SNE representation of embeddings obtained via neural network models and manually-designed features prepared as input for gradient boosting model.

\subsection{Classification of the day of week}

The first problem we consider for demonstration purposes is the classification of a day of week based on embeddings for this day.
The target accuracy is the percentage of correctly detected days of week. 

Best achieved results with its corresponded parameters are in Table~\ref{table:dow_prediction}. 
Neural network models work better, than baselines, making them a useful tool for processing of huge amount of data due to representation learning ability.
LSTM neural network performance is worse: 
its quality is similar to Gradient boosting with Word2vec input features. 
We suppose, that the reason is the high volatility of data, 
as we pick random subset of transactions for a day.
CNN on transaction sequences with length 2000 provides the best accuracy among all considered methods.

\begin{table}[h]
    \centering
    \begin{tabular}{ccc}
        \hline
         Model & Number of transactions & Accuracy\\
         & sampled every epoch & \\
        \hline
         %\multirow{}{}
         XGBoost & - & $0.45 \pm 0.01$ \\
         XGboost, Word2vec features & - & $0.55 \pm 0.02$ \\
        LSTM, triplet loss       & 500  & $0.54 \pm 0.02$ \\
        LSTM, cross-entropy loss & 500  & $0.58 \pm 0.02$ \\
        CNN, cross-entropy loss  & 2000 & $\mathbf{0.70} \pm \mathbf{0.02}$ \\
  
         \hline
    \end{tabular}
    \caption{Performance comparison for day-of-week prediction: CNN model shows the best results}
    \label{table:dow_prediction}
\end{table}

\begin{figure}[h]
\centering
\includegraphics[width=1\linewidth]{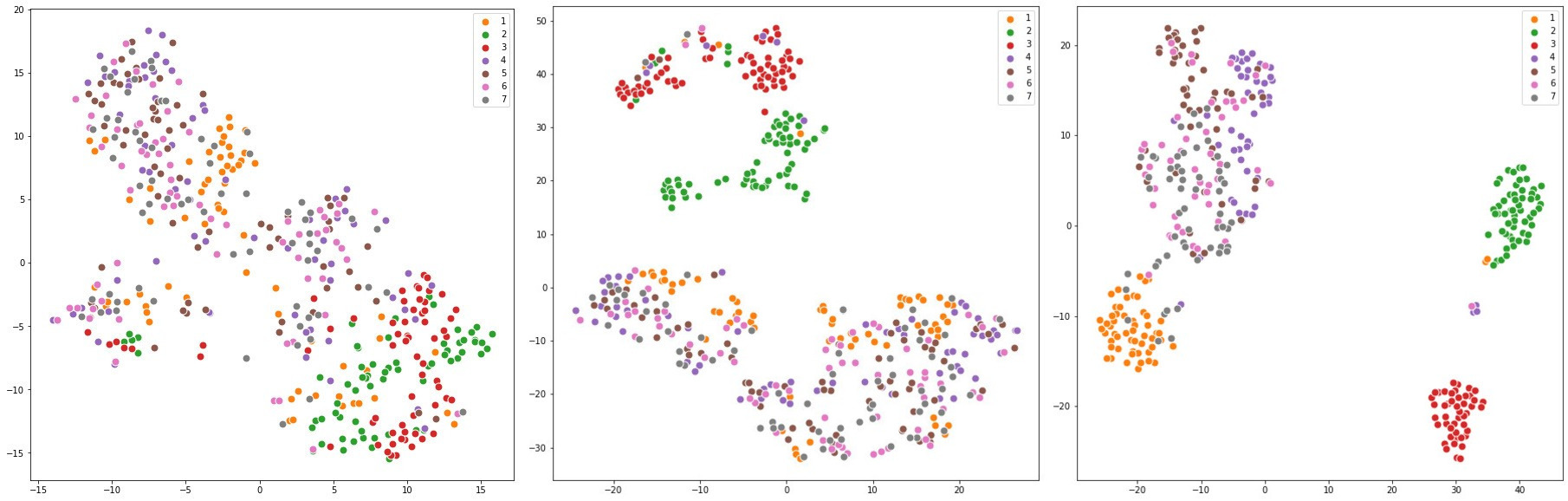}
 \caption{t-SNE embeddings computed with three models for days from model embeddings. Each color corresponds to a specific day-of-week. \emph{The left} figure presents result for the baseline, \emph{the middle one} presents result for the word2vec features, \emph{the right one} shows obtained embeddings for the CNN model with the best separation in the embeddings space.}
 \label{Figure:cnn}
\end{figure}

CNN provides the best accuracy, so we construct embeddings for it and for manually-selected features.
Comparison of the results obtained with simple baseline on statistical features, NLP and CNN models on raw data are in Figure~\ref{Figure:cnn}.
As expected from accuracy metric, CNN provides much better separation.

% The influence of $a$ parameter on embedding space was additionally investigated Fig  \ref{Figure:parameters_nlp1}. For $a$ less than $10^{-4}$ Saturday and Sunday are less separable Fig \ref{Figure:parameters_nlp2}. In general, value of $a$ doesn't influence of separability of weekdays/weekends.

% \begin{figure}[h]
% \centering
% \includegraphics[width=1\linewidth]{part1.jpg}
%  \caption{Choice of a parameter (from left to right): $10^{-2}$, $10^{-3}$, $10^{-4}$}
%  \label{Figure:parameters_nlp1}
% \end{figure}

% \begin{figure}[h]
% \centering
% \includegraphics[width=1\linewidth]{part2.jpg}
%  \caption{Choice of a parameter (from left to right): $10^{-5}$, $10^{-6}$, $10^{-7}$}
%  \label{Figure:parameters_nlp2}
% \end{figure}

\subsection{Regression of the default rate}

% \textit{CNN.} One-dimensional CNN is a popular choice for processing of long sequential data.
% We used raw transaction data with preprocessed categorical variables (MCC-code, transaction type, currency, country) via embedding layer with learnable embeddings matrix. We used Python language and a PyTorch package. Experiments were conducted with sequences of following lengths: 1000, 5000, 10000, 20000, 30000. Every epoch we resample sequence with the length of the specified above from dataset. The number of training epochs is $500$, initial learning rate is $\alpha = 0.1$ with step decay coefficient of 0.95 and step size of 5. We use standard Adam optimizer. Weights are initialized randomly. Embedding sizes are following $36$, $18$, $16$, $16$ for MCC-code,transaction type, country and currency respectively. Batch size is $4$. 

The two main classes of approaches for supervised regression problem for transaction sequence data are boosting on handcrafted features and neural-network-based models, such as LSTM and CNN. These approaches were implemented and compared with each other.

The present results are in Table~\ref{res2}.
To speak one at a time, the gradient boosting model performs worse than other models with negative validation $R^2$ score. 
It shows, how much more sophisticated and complex problem is the default rate prediction compare to day of week prediction. 
IndRNN model replacing its predecessor LSTM shows better results compare to the baseline with low positive $R^2$ score for test set.
The amount of available GPU memory limits the maximum sequences length for IndRNN model.
CNN model show adequate performance for default rate prediction problem. Experiments show, that performance of best model do not depend from sequence length.
Moreover, the CNN model trains 24 faster, than the IndRNN model. 

\begin{table}[h]
\begin{center}
\begin{tabular}{ccc}
\hline
% \textbf{Table}&\multicolumn{3}{|c|}{\textbf{Table Column Head}} \\
% \cline{2-4} 
\textbf{Model} & \textbf{\textit{N}}& \textbf{\textit{$R^2$}} \\
\hline
Baseline XGBoost & - & $-0.82 \pm {0.06}$  \\
IndRNN & 5000 & $0.19 \pm {0.01}$ \\
CNN & 10000 & $\mathbf{0.40} \pm \mathbf{0.02}$ \\
\hline
% \multicolumn{4}{l}{$^{\mathrm{a}}$Sample of a Table footnote.}
\end{tabular}
\caption{Performance comparison for the default rate prediction: CNN model shows the best results. We also show optimal processed sequence length $N$.}
\label{res2}
\end{center}
\end{table}

% Visual comparison of the prediction results obtained with simple baseline on statistical features, and CNN models presented on Figure \ref{Figure:default_rate}.
% We see, that CNN model can capture the trend in the data caused by a crisis at some point.

% \begin{figure}[h]
% \centering
% \includegraphics[width=1\linewidth]{figures/default_rate.png}
%  \caption{Results obtained with two models for the portfolio default rate prediction problem. The x- and y- axis is hidden, as they contain sensitive information. The scales of both axes are linear. The red vertical dashed line represents train/validation split. Top figure presents result for the gradient boosting baseline, the bottom one presents result for CNN model. Yellow CNN predictions better catch the trend compared to the baseline model.}
%  \label{Figure:default_rate}
% \end{figure}

\subsection{Stability of the embeddings}

Our approach as well produces embeddings for days in our data. 
As we claim in previous sections, such embeddings are useful, when predicting the portfolio default rate. 
We hope, that these embeddings can be useful in other problems, and are stable enough, so one can easily retrain them using new data and obtain similar results. The stability question arises in this work, because data is quite noisy and one should define how obtained embeddings are appropriate at all. What if they are random? Thats's why we define embeddings stability metrics, and then measure it for our embeddings.

\paragraph{Embeddings stability quality metric} 
Given a set \(S\) of \(N\) elements \(S=\{s_{1},s_{2},\ldots s_{N}\}\), consider two partitions of \(S\), namely \(U=\{U_{1},U_{2},\ldots ,U_{R}\}\) with \(R\) clusters, and \(V=\{V_{1},V_{2},\ldots ,V_{C}\}\) with \(C\) clusters.
Our goal is to measure correspondence between two partitions: how similar is $U$ to $V$.
We do it via Adjusted Mutual Information (AMI) score~\cite{vinh2010information}.

The entropy of the partition \(U\) is \(H(U)\), the entropy of the partition \(V\) is \(H(V)\). \(MI(U, V)\) is the mutual information between two partitions. Then, Adjusted Mutual Information (AMI) is calculated in the following way:
\[
AMI(U, V) = \frac{MI(U, V) - \mathbb{E}\{MI(U, V)\}}{\max\{H(U), H(V)\} - \mathbb{E}\{MI(U, V)\}},
\]
where $\mathbb{E}$ corresponds to expected values for a random partition.
AMI is an adjusted partition quality metric, such that for a random partitions $U$ and $V$ it will have value $0$, and the ideal correspondence will have AMI value $1$ with bigger values are better.

\paragraph{AMI estimation}
To measure stability of embeddings we estimate AMI for our embeddings in the following way:
\begin{itemize}
\item Sample transactions 10 times from original dataset.
\item Obtain 10 sets of embeddings with a separate training of a suggested algorithm.
\item Cluster them via K-means clustering with 7 clusters.
\item Calculate AMI for obtained between each pair from 45 possible pairs.
\item Average 45 obtained values.
\end{itemize}
Mean AMI is $0.33$ versus $0.00$ for random embeddings.
Thus, we conclude that our embedding vectors are stable enough to be used in other problem and with new data.

\section{Conclusion}

The current work investigates the problem of macroeconomic indexes prediction based on financial transactions data. The challenge here is the tremendous amount of available semi-structured transactions: millions of clients commit transactions on a specific day. 

We propose an approach based on a sampling of transaction sequences and processing these sequences via a neural network model. The evidence of the superiority of our approach comes from results for two problems: a portfolio of credits default rate prediction and a day-of-week prediction.

We consider two main directions for constructing such models: classic feature-based machine learning and representation learning via neural networks. For the classic gradient boosting model, we use hand-crafted features and word2vec features. For neural-network-based models LSTM and CNN, we use sampling and end2end learning for sampled sequences of transactions. CNN provides the best accuracy of 0.70 for day-of-week prediction. According to the results for this problem, we picked up two models: boosting on hand-crafted features and end2end CNN on raw data to continue with them for the default rate prediction task. For the second problem, the best $R^2 = 0.4$ is again for CNN. So, it seems reasonable to apply CNN for such long sequences with hundreds of thousands of transactions. 

Our neural network predicts the default rate with high quality --- using only raw transactions data. The reasons are a new sampling scheme and an end2end training procedure. 
The demonstrated approach shows the correlation between the people's transaction activity and the portfolio default rate and predicts the future default rate. Bank employees can use obtained embeddings to solve other downstream tasks making them universal indicators of the current macroeconomic state. 

\section*{Acknowledgments}

The work is supported  by the Russian Science Foundation (project 21-11-00373). 

\bibliography{references.bib}
\bibliographystyle{IEEEtran}

\end{document}